\documentclass[aps,prb,preprint,showpacs,amsfonts]{revtex4}

\usepackage{epsfig}
\usepackage{amsmath}
\usepackage{amssymb}

\newcommand{\beq}{\begin{equation}}
\newcommand{\eeq}{\end{equation}}
\newcommand{\beqarray}{\begin{eqnarray}}
\newcommand{\eeqarray}{\end{eqnarray}}
\newcommand{\half}{\ensuremath{\tfrac{1}{2}}}
\newcommand{\Hc}{\ensuremath{\mbox{H.c.}}} 
\newcommand{\Ham}[1][]{\ensuremath{{\cal{H}}_{\text{\tiny{#1}}}}} 
\newcommand{\eq}[1]{Eq.~(\ref{#1})} 
\newcommand{\fig}[1]{Fig.~\ref{#1}} 
\newcommand{\Sec}[1]{Sec.~\ref{#1}} 
\newcommand{\Ref}[1]{Ref.~\onlinecite{#1}} 

\begin{document}

\title{Slave boson theory of the extended Falicov-Kimball model}
\author{P. M. R. Brydon} 
\affiliation{Max-Planck-Institut f\"{u}r Festk\"{o}rperforschung,
  Heisenbergstr. 1, 70569 Stuttgart, Germany}

\date{\today}

\begin{abstract}
The extended Falicov-Kimball model, with both an on-site hybridization
potential and dispersive narrow band, is examined within the saddle-point
approximation to the Kotliar-Ruckenstein slave boson theory. We first set the
hybridization potential to zero and find that the phase diagram depends
strongly upon the orbital structure: for degenerate orbitals,
a correlated-insulating state is found at sufficiently strong interaction
strengths, whereas 
a finite orbital energy difference can lead to discontinuous valence
transitions. The obtained phase diagram is very sensitive to the presence of a
finite hybridization potential. As in Hartree-Fock theory, we find an
enhancement of the hybridization by the inter-orbital Coulomb repulsion. The
more precise treatment of correlation effects, however, leads to large
deviations from the Hartree-Fock results.
In the limit of vanishing hybridization an excitonic
insulator state is only found when the orbitals are degenerate, which
restricts this phase to a much smaller parameter space than
in other available mean-field theories.
\end{abstract}

\maketitle

\section{Introduction}

The Falicov-Kimball model (FKM) was one of the first theoretical attempts to
explain valence transitions in mixed-valence 
systems such as SmB$_6$ and Ce.~\cite{FKMorig} In its original 
form, the model describes a spinless fermion system with conduction ($c$)
electrons interacting 
via on-site Coulomb repulsion $G$ with narrow band ($f$) electrons of
orbital energy $\epsilon_{f}$. Within a mean-field theory, it was found that
varying 
$G$ or $\epsilon_{f}$ can produce both continuous and
discontinuous changes in the distribution of electrons between these two
orbitals, i.e. valence transitions. The FKM is nevertheless not a good model
of the mixed-valence state as the entirely localized nature of the $f$
electrons is 
unrealistic.~\cite{LRP81} The ``classical'' nature of the $f$ electrons was
subsequently exploited in the reinterpretation of the FKM as a model
of charge order in binary alloys.~\cite{BinaryAlloy}

The central idea behind the FKM, that the inter-orbital Coulomb repulsion $G$
could be the origin of discontinuous valence transitions, was revisited by
several groups who modified the model to account for the quantum nature of the
$f$ electrons.~\cite{FKMhyb,Khomskii,Leder78} The so-called extended
Falicov-Kimball model (EFKM) allows for the partial delocalization of the $f$
electrons due to the overlap of the orbital wave functions by including a
$c$-$f$ hybridization potential $V$. Although mean-field impurity models
suggested that discontinuous valence transitions were 
possible in the EFKM,~\cite{FKMhyb,Khomskii} weak-coupling studies of the
periodic limit found only continuous changes in the orbital
populations.~\cite{Leder78,BC82} 

An interesting aspect of the Hartree-Fock (HF) solution of the EFKM with
on-site hybridization is the existence of an excitonic insulator (EI) phase in
the limit $V\rightarrow0$.~\cite{Khomskii} The EI phase is characterized by a
spontaneous hybridization between the $c$ and $f$ bands due to the presence of
a non-zero polarization or excitonic average
$\langle{c^{\dagger}f}\rangle$. This is interpreted as indicating the
spontaneous pairing of $c$ electrons with $f$ holes, forming an excitonic
condensate. Introduced independently by Keldysh and Kopaev and des
Cloizeaux,~\cite{KK65,dC65} the EI is
an important concept in the study of semimetal-semiconductor transitions. 
A spontaneous hybridization between the nested portions of the Fermi surface
has proved a particularly useful description of the the spin density-wave
phase of Cr-based alloys;~\cite{Chromium} a similar scenario has also recently
been proposed as an explanation for the orbital ordering in
LaMnO$_3$.~\cite{EK05} Unambiguous examples of the EI phase remain rare,
however, with only 
two likely candidates, the alloys Sm$_{0.9}$La$_{0.1}$S and
TmSe$_{0.45}$Te$_{0.55}$, thus far discovered.~\cite{EIexp} This indicates
that the conditions for an EI phase must be significantly more restrictive
than those encountered in the usual phenomenological Hamiltonian description.

The EFKM is the only ``standard'' tight-binding model of correlated
electron systems that has been claimed to display an EI phase.~\cite{Khomskii}
It has attracted much attention due to the proposal by Portengen
\emph{et al.} that the spontaneous excitonic average in the EFKM could be
interpreted as evidence of electronic
ferroelectricity.~\cite{POS96} Although a variety of more
sophisticated treatments~\cite{Farky,Zlatic01} or more general mean-field
theories~\cite{C99,BZGB05} have failed to find the EI phase, the presence
of a finite $f$-electron hopping can stabilize the EI state
in the strong-coupling regime.~\cite{Batista,SC04} Furthermore, it seems
likely that in the EFKM with $V\neq0$ the inter-orbital Coulomb interaction
will induce a large ``excitonic'' renormalization of the bare on-site
hybridization potential.~\cite{C99,BZGB05} 

The continuing controversy regarding the EI phase in the EFKM and the larger
question about the rarity of EI phases motivates us to study
the EFKM using a more advanced analytical technique than the weak-coupling
methods hitherto employed. 
A particularly useful analytic approach for obtaining the ground state
properties of strongly-correlated lattice models is the slave boson (SB)
mean-field theory developed by Kotliar and Ruckenstein.~\cite{KR86} 
This is superior to HF mean-field theory as it accounts for the
renormalization of the quasiparticle weight by the interactions, similar to
a Fermi-liquid description.
It is also of interest to study the SB solution of the EFKM for the possible
application to multi-band Hubbard models:~\cite{SBmultiband} in general, the
presence of an on-site hybridization makes it impossible to apply the usual SB
formalism as the atomic Hamiltonian cannot then be written only in terms of
density operators. 
Although generalizations of the SB mean-field theory have been developed to
cope with these difficulties,~\cite{SBgeneralization} the effect of
inter-orbital interactions on the hybridization is still poorly understood. We
can obtain some insight into this situation by studying the EFKM with a finite
hybridization, as this can be treated within the usual SB formulation

In this paper we examine the EFKM at zero temperature ($T=0$) and half-filling
using the Kotliar-Ruckenstein SB
theory. In~\Sec{sec:SBtheory} we outline the construction of the mean-field SB
Hamiltonian as well as reviewing the usual HF solution. In both cases, we
consider only uniform ground states. 
The results are presented in~\Sec{sec:results}. 
The solution of the $V=0$ system (\Sec{subsec:Veq0}) is found to be very
sensitive to the orbital structure. For degenerate $c$ and $f$ orbitals, a
Brinkman-Rice-like insulating state is found at sufficiently large interaction
strength; for non-degenerate orbitals, discontinuous valence transitions can
be found. The orbital structure also determines the behaviour of the EFKM with
$V\neq0$ and $t_{ff}\leq0$ (\Sec{ss:Vne0tle0}): for degenerate orbitals, the 
SB solution resembles closely the predictions of the HF theory; for
non-degenerate orbitals, the more accurate treatment of correlation effects in
the SB theory produces strong deviations from the HF results. A first-order
metal-insulator transition is found with increasing interaction strength for
$t_{ff}>0$ (~\Sec{ss:Vne0tgt0}), in contrast to the second-order transitions
found in HF theory. Within the SB treatment of the model, the
EI phase is only possible when the $c$ and $f$ orbitals are degenerate
(\Sec{sss:EI}). We conclude in~\Sec{sec:concl} with a summary of
our results and outlook for further work.

\section{Slave-Boson Hamiltonian} \label{sec:SBtheory}

\begin{figure}
\includegraphics[width=9.5cm]{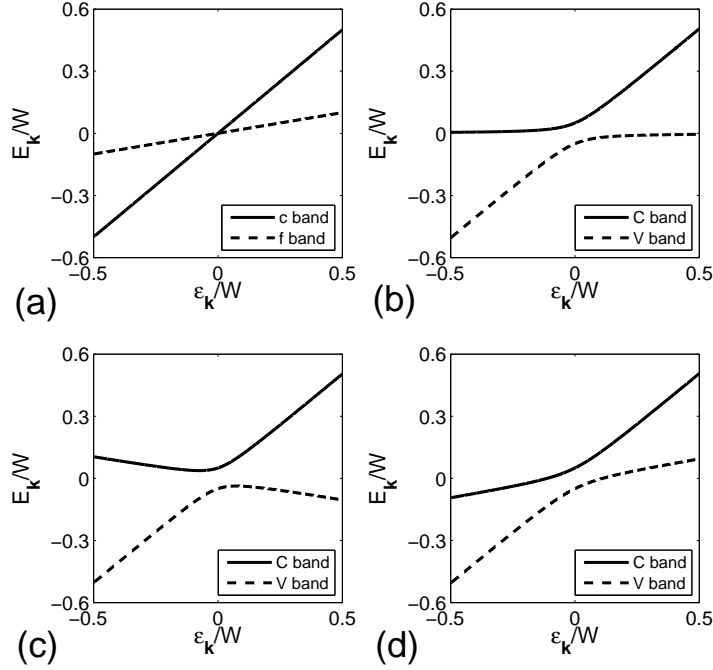} 
\caption{\label{bands} Different band scenarios for the non-interacting
  EFKM with $\epsilon_{f}=0$. $W$ 
  is the bandwidth of the bare $c$ electron band. (a) $c$  and $f$ electron
  bands for $t_{ff}=0.2$, $V=0.0$; (b) C  and V bands for
  $t_{ff}=0.0$, $V=0.05W$; (c) C  and V bands for
  $t_{ff}=-0.2$, $V=0.05W$; (d) C  and V bands for
  $t_{ff}=0.2$,  $V=0.05W$.}  
\end{figure}

The Hamiltonian for the EFKM is written
\beq
\Ham[EFKM] = \sum_{\bf{k}}\epsilon_{\bf{k}}c^{\dagger}_{\bf{k}}c^{}_{\bf{k}} +
\sum_{\bf{k}}(t_{ff}\epsilon_{\bf{k}} +
\epsilon_{f})f^{\dagger}_{\bf{k}}f^{}_{\bf{k}} + V\sum_{\bf{k}}\left\{c^{\dagger}_{\bf{k}}f^{}_{\bf{k}} + \Hc\right\} +
G\sum_{j}n^{c}_{j}n^{f}_{j} \label{eq:HamEFKM1}
\eeq
where $c^{}_{\bf{k}}$ ($c^{}_{j}$) and $f^{}_{\bf{k}}$ ($f^{}_{j}$) are the
annihilation operators in momentum (real) space for the $c$ and $f$ electrons
respectively. The $c$ electron dispersion is $\epsilon_{\bf{k}}$; the
$f$ electron 
dispersion is assumed to be a multiple $|t_{ff}|<1$ of the $c$-electron
dispersion. In this work we consider hole-like ($t_{ff}<0$), dispersionless
($t_{ff}=0$) and electron-like ($t_{ff}>0$) $f$ electron bands. For $G=0$, we
illustrate the resulting band dispersions as a function of $\epsilon_{\bf{k}}$
in~\fig{bands}. For $V\neq0$, the 
bands are of mixed $c$- and $f$-character: we refer to the upper and lower
bands as the conduction (C) and valence (V) bands respectively.
Note that for $t_{ff}\leq0$ the non-interacting ground state is insulating for
arbitrarily small $V$, while it is metallic for sufficiently large $t_{ff}>0$.

In the absence of a hybridization potential between the $c$- and $f$-electron
states we may immediately apply the SB technique. In the case
when $V\neq0$, however, the SB technique cannot be
straight-forwardly 
applied as the atomic Hamiltonian (obtained by neglecting all kinetic terms
in~\eq{eq:HamEFKM1}) is not diagonal  
in the electron occupation operators.~\cite{SBgeneralization}
To proceed, we re-write the system in terms of a diagonal on-site basis
\beq
b^{}_{j} = \alpha{c^{}_{j}} + \beta{f^{}_{j}} \qquad a^{}_{j} = \beta{c^{}_{j}} - \alpha{f^{}_{j}}
\label{eq:abbasis}
\eeq
where
\beqarray
\alpha &=& \frac{\mbox{sgn}(\epsilon_{f}+0^+)}{\sqrt{2}}\left[{1 + \sqrt{1 -
    \frac{4V^2}{4V^2+\epsilon_{f}^2}}}\right]^{1/2} \\
\beta &=& \frac{1}{\sqrt{2}}\left[{1 -
  \sqrt{1 - \frac{4V^2}{4V^2+\epsilon_{f}^2}}}\right]^{1/2}
\eeqarray
Note that $\mbox{sgn}(x+0^+)=1 (-1)$ for $x\geq0 (x<0)$.
In terms of the $a$ and $b$ operators we can hence re-write the Hamiltonian
\beqarray
\Ham[EFKM] & = &
\sum_{\bf{k}}\epsilon_{\bf{k}}\left\{(\alpha^2 +
  t_{ff}\beta^2){b}^{\dagger}_{\bf{k}}{b}^{}_{\bf{k}} 
+
(\beta^2+t_{ff}\alpha^2){a}^{\dagger}_{\bf{k}}{a}^{}_{\bf{k}} 
+ (1-t_{ff})\alpha\beta\left[{a}^{\dagger}_{\bf{k}}{b}^{}_{\bf{k}}
+ {b}^{\dagger}_{\bf{k}}{a}^{}_{\bf{k}}\right]\right\}
\notag \\
&& +
\sum_{j}\epsilon_a{a}^{\dagger}_{j}{a}^{}_{j}
+
\sum_{j}\epsilon_b{b}^{\dagger}_{j}{b}^{}_{j}
 + G\sum_{j}n^{a}_{j}n^{b}_{j} 
\eeqarray
where $\epsilon_{b} = \epsilon_{f}\beta^2 - 2\alpha\beta{V}$ and $\epsilon_{a}
= \epsilon_{f}\alpha^2 + 2\alpha\beta{V}$. 

We adopt the Kotliar-Ruckenstein SB theory by introducing the
auxiliary bosonic fields $e_{j}$, $s_{aj}$, $s_{bj}$ and $d_{j}$ which
respectively destroy the empty, singly-occupied $a$ orbital,
singly-occupied $b$ orbital and doubly-occupied atomic configurations at site
$j$. The fermionic Hamiltonian is then written in terms of quasi-fermions
$\widetilde{a}$ and $\widetilde{b}$ using the identification 
\beqarray
a^{}_{j} & = & z_{aj}\widetilde{a}^{}_{j} \\
b^{}_{j} & = & z_{bj}\widetilde{b}^{}_{j}
\eeqarray
where
\beq
z_{a(b)j} = (1-d^{\dagger}_{j}d^{}_{j}-s^{\dagger}_{a(b)j}s^{}_{a(b)j})^{-1/2}(s^{\dagger}_{b(a)j}d^{}_{j}+e^{\dagger}_{j}s^{}_{a(b)j})(1-s^{\dagger}_{b(a)j}s^{}_{b(a)j}-e^{\dagger}_{j}e^{}_{j})^{-1/2}.
\eeq
The physical interpretation of the bosonic fields implies that the following
equations are satisfied at each site
\beqarray
1 & = & e^{\dagger}_{j}e^{}_{j} + s^{\dagger}_{aj}s^{}_{aj} +
s^{\dagger}_{bj}s^{}_{bj} + d^{\dagger}_{j}d^{}_{j} \\
\widetilde{a}^{\dagger}_{j}\widetilde{a}^{}_{j} & = & s^{\dagger}_{aj}s^{}_{aj} +
d^{\dagger}_{j}d^{}_{j} \\
\widetilde{b}^{\dagger}_{j}\widetilde{b}^{}_{j} & = & s^{\dagger}_{bj}s^{}_{bj} +
d^{\dagger}_{j}d^{}_{j}.
\eeqarray
These constraints are respectively enforced by the constraint fields
$\lambda_{j}$, $\Lambda_{aj}$ and $\Lambda_{bj}$, which enter as Lagrangian
multipliers. 

A mean-field theory is constructed by replacing the boson and
constraint fields by spatially-uniform time-invariant fields,
i.e. $e_{j}\rightarrow{e}$, $s_{bj}\rightarrow{s_b}$, etc.. This yields the 
Hamiltonian 
\beqarray
\Ham[SB] & = &
\sum_{\bf{k}}\epsilon_{\bf{k}}\left\{z_b^2(\alpha^2 +
  t_{ff}\beta^2)\widetilde{b}^{\dagger}_{\bf{k}}\widetilde{b}^{}_{\bf{k}} 
+
z_a^2(\beta^2+t_{ff}\alpha^2)\widetilde{a}^{\dagger}_{\bf{k}}\widetilde{a}^{}_{\bf{k}} 
+ z_{a}z_{b}(1-t_{ff})\alpha\beta\left[\widetilde{a}^{\dagger}_{\bf{k}}\widetilde{b}^{}_{\bf{k}}
+ \widetilde{b}^{\dagger}_{\bf{k}}\widetilde{a}^{}_{\bf{k}}\right]\right\}
\notag \\
&& +
\sum_{\bf{k}}(\epsilon_a+\Lambda_a)\widetilde{a}^{\dagger}_{\bf{k}}\widetilde{a}^{}_{\bf{k}}
+
\sum_{\bf{k}}(\epsilon_b+\Lambda_b)\widetilde{b}^{\dagger}_{\bf{k}}\widetilde{b}^{}_{\bf{k}}
\notag \\
& & + NGd^{2} - N\lambda(e^{2}+s_{a}^{2} + s_{b}^{2} + d^{2}-1) -
  N\Lambda_{a}(s_{a}^2+d^{2}) - N\Lambda_{b}(s_{b}^2+d^{2}) \label{eq:sbHam1}
\eeqarray
where 
\beq
z_{a(b)} = (1-d^2-s_{a(b)}^2)^{-1/2}(ds_{b(a)}+es_{a(b)})(1-s_{b(a)}^{2}-e^{2})^{-1/2} 
\eeq
are the band-renormalization factors. We work throughout at half-filling
\beq
1 =
\frac{1}{N}\sum_{j}\left\{\langle\widetilde{a}^{\dagger}_{j}\widetilde{a}^{}_{j}\rangle
+
\langle\widetilde{b}^{\dagger}_{j}\widetilde{b}^{}_{j}\rangle\right\},
\label{eq:halffill} 
\eeq
as in this limit it may be explicitly demonstrated by extremization of
the free energy that $z_{a}=z_{b}=z$. 

The quasi-fermion component of the Hamiltonian~\eq{eq:sbHam1} can be
straight-forwardly diagonalized. For finite $c$-$f$ hybridization, we have the
quasi-fermion C and V bands 
\beq
E^{\scriptstyle{\text{C(V)}}}_{\bf{k}} = \frac{1}{2}\left\{(1+t_{ff})z^2\epsilon_{\bf{k}} +
  \widetilde{\epsilon}_{a} + \widetilde{\epsilon}_{b} + (-)
  \sqrt{[(1-t_{ff})z^2\epsilon_{\bf{k}} + \widetilde{\epsilon}_{a} -
    \widetilde{\epsilon}_{b}]^2 + 4\widetilde{V}^{2}}\right\}
\eeq
where
\beqarray
\widetilde{\epsilon}_{a} & = & \alpha^2(\epsilon_{b} + \Lambda_{b}) +
\beta^2(\epsilon_{a} + \Lambda_{a}) \\
\widetilde{\epsilon}_{b} & = & \beta^2(\epsilon_{b} + \Lambda_{b}) +
\alpha^2(\epsilon_{a} + \Lambda_{a}) \\
\widetilde{V} & = & \alpha\beta(\epsilon_{a} + \Lambda_{a} - \epsilon_{b} -
\Lambda_{b}). \label{eq:sbhyb}
\eeqarray
Note the renormalization of the hybridization by the constraint fields: this
is the equivalent of the excitonic enhancement seen in HF studies. Of
particular importance then is the so-called excitonic average, defined 
\beq
\Delta = 
\frac{1}{N}\sum_{\bf{k}}\langle{c^{\dagger}_{\bf{k}}f^{}_{\bf{k}}}\rangle
=
\frac{\alpha\beta}{N}\sum_{\bf{k}}\left\{\langle{a^{\dagger}_{\bf{k}}a^{}_{\bf{k}}}\rangle  
- \langle{b^{\dagger}_{\bf{k}}b^{}_{\bf{k}}}\rangle\right\}. \label{eq:sbDelta}
\eeq
If $\Delta$ remains finite as $V\rightarrow0$, the system has an instability
towards the EI phase. 

The familiar SB self-consistency conditions are obtained by
minimizing the free energy with respect to the SB fields 
while maximizing with respect to the constraint fields, the so-called
saddle-point approximation. The free energy may be calculated
analytically in the case of a rectangular density of states (DOS)
\beq
\rho(\omega) = 
\begin{cases}
1/W, & |\omega|<W/2 \\
0, & |\omega|>W/2
\end{cases} \label{eq:DOS}
\eeq
where the DOS of the bare $c$ and $f$ bands is respectively $\rho_{c}(\omega)
= \rho(\omega)$ and $\rho_{f}(\omega) =
|t_{ff}|^{-1}\rho(t_{ff}^{-1}[\omega-\epsilon_{f}])$. Although the details of
the self-consistent solutions will change upon adopting a more realistic
band-structure, the rectangular DOS is convenient for studying the generic
behaviour of the model. 

\subsection{Hartree-Fock Theory}

For comparison, we briefly discuss the usual HF solution of the
EFKM.~\cite{Leder78} An effective single-particle Hamiltonian may be derived
from~\eq{eq:HamEFKM1} by decoupling the interaction term 
\beqarray
G\sum_{j}n^{c}_{j}n^{f}_{j} &\approx& Gn_c\sum_{j}n^{f}_j +
Gn_{f}\sum_{j}n^{c}_{j} - G\Delta\sum_{j}\left\{c^{\dagger}_{j}f^{}_{j} +
  \Hc\right\} \notag \\
&& - NGn_{c}n_{f} + NG\Delta^2 \label{eq:HFdecomp}
\eeqarray
where the HF variational parameters $n_{c}$, $n_{f}$ and $\Delta$ are
respectively the $c$ electron concentration, the $f$ electron concentration
and the excitonic average defined
in~\eq{eq:sbDelta}. Substituting~\eq{eq:HFdecomp} 
into~\eq{eq:HamEFKM1} we obtain the mean-field Hamiltonian 
\beqarray
\Ham[HF] & = & \sum_{\bf{k}}(\epsilon_{\bf{k}} +
\widetilde{\epsilon}_{c})c^{\dagger}_{\bf{k}}c^{}_{\bf{k}} + 
\sum_{\bf{k}}(t_{ff}\epsilon_{\bf{k}} +
\widetilde{\epsilon}_{f})f^{\dagger}_{\bf{k}}f^{}_{\bf{k}} +
\widetilde{V}\sum_{\bf{k}}\left\{c^{\dagger}_{\bf{k}}f^{}_{\bf{k}} +
  \Hc\right\} \notag \\ 
&& - NGn_{c}n_{f} + NG\Delta^2 \label{eq:HFHam}
\eeqarray
where $\widetilde{\epsilon}_{c} = Gn_{f}$, $\widetilde{\epsilon}_{f} =
\epsilon_{f} + Gn_{c}$ and $\widetilde{V} = V - G\Delta$. The self-consistency
equations for the HF parameters are easily found by diagonalization
of the Hamiltonian~\eq{eq:HFHam}, see for example~\Ref{Leder78}. As for the SB
results, we calculate the self-consistency equations analytically using the
rectangular DOS~\eq{eq:DOS}.

Since the SB parameters are related to the concentration of sites in a given
orbital configuration, we can also obtain HF values of these parameters
directly from the HF wavefunction $|\Psi_{HF}\rangle$. Explicitly, we have
\beqarray
d^2_{HF} & = &
\frac{1}{N}\sum_{j}\langle\Psi_{HF}|n^{c}_{j}n^{f}_{j}|\Psi_{HF}\rangle \notag
\\
& = & n_{c}(1-n_c) - \Delta^2 \label{eq:dHF} \\
s^{2}_{bHF} & = & \frac{1}{N}\sum_{j}\langle\Psi_{HF}|\alpha^2n^{c}_{j} +
\beta^2n^{f}_{j} +
\alpha\beta\{c^{\dagger}_{j}f^{}_{j}+f^{\dagger}_{j}c^{}_{j}\}|\Psi_{HF}\rangle
- d^{2}_{HF} \notag \\
& = & \beta^2 + (\alpha^2-\beta^2)n_{c} - n_{c}(1-n_{c}) + 2\alpha\beta\Delta
+ \Delta^2 \label{eq:sbHF} \\
s^{2}_{aHF} & = & \frac{1}{N}\sum_{j}\langle\Psi_{HF}|\beta^2n^{c}_{j} +
\alpha^2n^{f}_{j} -
\alpha\beta\{c^{\dagger}_{j}f^{}_{j}+f^{\dagger}_{j}c^{}_{j}\}|\Psi_{HF}\rangle
- d^{2}_{HF} \notag \\
& = & \alpha^2 + (\beta^2-\alpha^2)n_{c} - n_{c}(1-n_c) - 2\alpha\beta\Delta +
\Delta^2 \label{eq:saHF} \\
e^2_{HF} & = & 1 - s^2_{aHF} - s^2_{bHF} - d^2_{HF} \notag \\
& = & n_{c}(1-n_c) - \Delta^2 \label{eq:eHF}
\eeqarray

\section{Results} \label{sec:results}

As we work throughout at $T=0$ and fixed particle number, we perform the
extremization upon the ground state energy per site
$E=\langle{\Ham[SB]}\rangle/N$. This is calculated analytically using the
density of states~\eq{eq:DOS}. The physical values of the SB and constraint
fields are then obtained by determining the saddle point of the ground state
energy, which requires that we solve the equations 
\beq
\frac{\partial{E}}{\partial{e}} = \frac{\partial{E}}{\partial{s_{a}}} =
\frac{\partial{E}}{\partial{s_{b}}} = \frac{\partial{E}}{\partial{d}} =
\frac{\partial{E}}{\partial{\lambda}} =
\frac{\partial{E}}{\partial{\Lambda_{a}}} =
\frac{\partial{E}}{\partial{\Lambda_{b}}} = 0. \label{eq:saddle_point}
\eeq
We employ a multi-dimensional Newton-Raphson technique to
solve~\eq{eq:saddle_point}. To obtain the HF results we iterate the
self-consistency equations until a desired accuracy is obtained.

\subsection{$V=0$, $t_{ff}\neq0$} \label{subsec:Veq0}

\begin{figure}
\includegraphics[width=9.5cm]{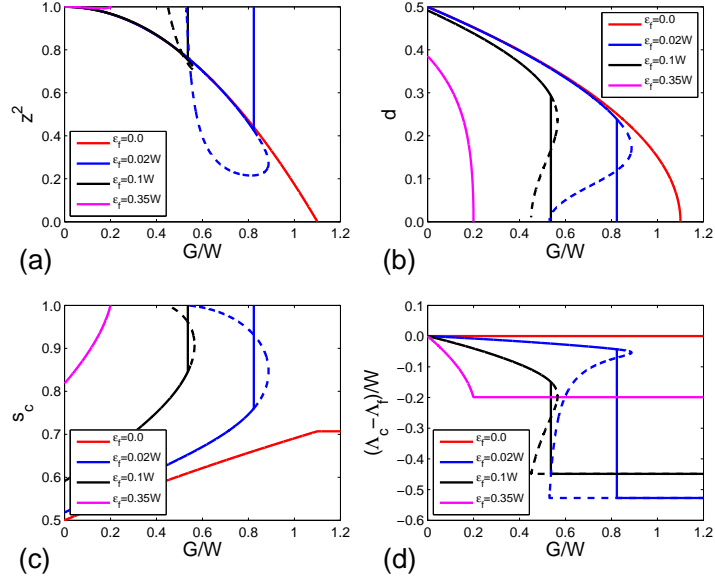} 
\caption{\label{sbfields_Veq0} (color online) Variation of the slave boson (SB)
  parameters with $G$ for $t_{ff}=0.1$, $V=0$ and various values of
  $\epsilon_{f}$. The solid lines indicate the ground state solution; the
  dashed lines give the metastable solutions. (a) Band renormalization factor
  $z^2$; (b) SB field $d$; (c) SB field $s_{c}$; (d)
  constraint field $\Lambda_{c}-\Lambda_{f}$.}   
\end{figure} 

We begin by examining the EFKM without hybridization. Mathematically, this
limit is very closely related to the uniform Gutzwiller and SB solutions of the
Hubbard model in a magnetic field:~\cite{KR86,Vollhardt84} identifying the
longitudinal 
magnetic field with $\epsilon_{f}$, the expression for $E$ in the two models
is of the same form, although the effective bandwidth of the EFKM is smaller by
a factor $(1+|t_{ff}|)/2$. This similarity implies that within the SB
approximation and assuming uniform ground states, the behaviour of the
EFKM is a charge analogue of the paramagnetic 
Hubbard model. In particular, there is a localization transition at
$\epsilon_{f}=0$, and at sufficiently small $\epsilon_{f}\neq0$ a first-order
valence transition occurs. This is displayed in our
plots of the SB fields in~\fig{sbfields_Veq0}. We do not discuss the
HF predictions for the EFKM with $V=0$ as this only involves the
renormalization of the orbital energies, see~\eq{eq:HFHam}.

At $\epsilon_{f}=0$ there is a second-order transition into a
Brinkman-Rice-like correlated-insulator (CI) phase at $G/W=1+|t_{ff}|$. 
This is a 
localization transition, as the band-renormalization factor $z^2$ vanishes in
the CI phase. In 
the CI phase, every site is singly-occupied with equal probability by either a
$c$ or $f$ electron, reflected in the limiting values 
$s_{c}=s_{f}=1/\sqrt{2}$ in~\fig{sbfields_Veq0}(c) and $d=0$
in~\fig{sbfields_Veq0}(b). As implied by the 
equality of $c$ and $f$ electron populations, the difference
between the effective $c$ and $f$ energy levels, $\Lambda_{c}-\Lambda_{f}$, is
zero for all $G$.

For any finite $\epsilon_{f}$ the high-$G$ state is an integer
valence state with filled $c$ band ($f$ band) for $\epsilon_{f}>0$
($\epsilon_{f}<0$). This filled-band (FB) state is not localized, so
we have $z=1$ [\fig{sbfields_Veq0}(a)]; at $\epsilon_{f}>0$ it is also
characterized by $s_{c}=1$ and 
$d=e=s_{f}=0$ [\fig{sbfields_Veq0}(b,c)]. As $G$ is increased,
$\Lambda_{c}-\Lambda_{f}$ decreases to the 
fixed value $\epsilon_{f} - \half(1+|t_{ff}|)W$ in the FB phase: this raises
the effective $f$ level so that the bottom of the $f$ band just touches the
top of the $c$ band. 

\begin{figure}
\includegraphics[width=9.5cm]{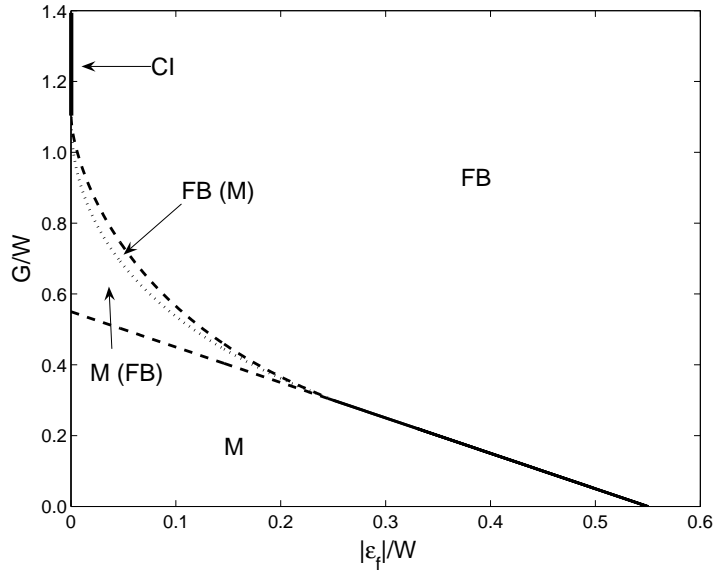} 
\caption{\label{PD_Veq0} Phase diagram for the EFKM in the $G$-$\epsilon_{f}$
  plane. We have $|t_{ff}|=0.1$ and $V=0$. At $\epsilon_{f}=0$, we have a
  second-order transition between the metal (M) phase and the correlated
  insulator (CI) phase. The limits
  of metastability of the filled band (FB) and M phases are given by the
  dotted lines; along 
  the dashed line the two phases have equal energy. Where metastable state
  exists, it follows after the stable state in brackets. The solid line for
  $|\epsilon_{f}|\gtrsim0.25W$ gives the second-order transition 
  between the M and FB phases.}   
\end{figure}

The transition into the FB state from the metal is an example of a valence
transition, as it is a transition from a state with mixed valence
(i.e. non-zero $c$ and $f$ populations) into a state with integer occupation
of the orbitals. This  
transition is 
qualitatively different at small and large $|\epsilon_{f}|$. For small
$|\epsilon_{f}|$, the FB state is reached from the low-$G$ metallic (M) phase
by a first-order transition, which is the charge analogue of the metamagnetic
transition in the SB treatment of the Hubbard model. 
Within the region where both FB and M solutions of~\eq{eq:saddle_point} are
possible, the stable ground state is defined to be the one with the lower
energy. Multiple solutions are only found within the region bounded between
the dashed lines and $\epsilon_{f}=0$ in~\fig{PD_Veq0}, with the metastable
state indicated within the brackets following the stable state. The FB and M
phases have equal energy along the dotted line.
As $|\epsilon_{f}|$ is increased, the boundaries of metastability of the FB
and M phases converge together at 
$|\epsilon_{f}|\approx0.25W$, beyond which we find a second-order transition
between the M and FB phases. As $|\epsilon_{f}|$ is further increased, the
line of second order transitions intersects the line $G=0$ at
$|\epsilon_{f}|=\half(1+t_{ff})W$: for
$|\epsilon_{f}|>\half(1+t_{ff})W$ there is no overlap between the $c$ and
$f$ bands in the $G=0$ limit and so the system is always in the FB state. 

\subsection{$V\neq0$, $t_{ff}\leq0$} \label{ss:Vne0tle0}

The presence of a finite hybridization potential has a dramatic effect upon
the EFKM with $t_{ff}\leq0$, as the system is then in an insulating state at
$G=0$ as illustrated in~\fig{bands}(b,c). We find, however, that the cases
$\epsilon_{f}=0$ and $\epsilon_{f}\neq0$ are distinguished by very different
behaviour in the $V\rightarrow0$ limit. 

\begin{figure}
\includegraphics[width=9.5cm]{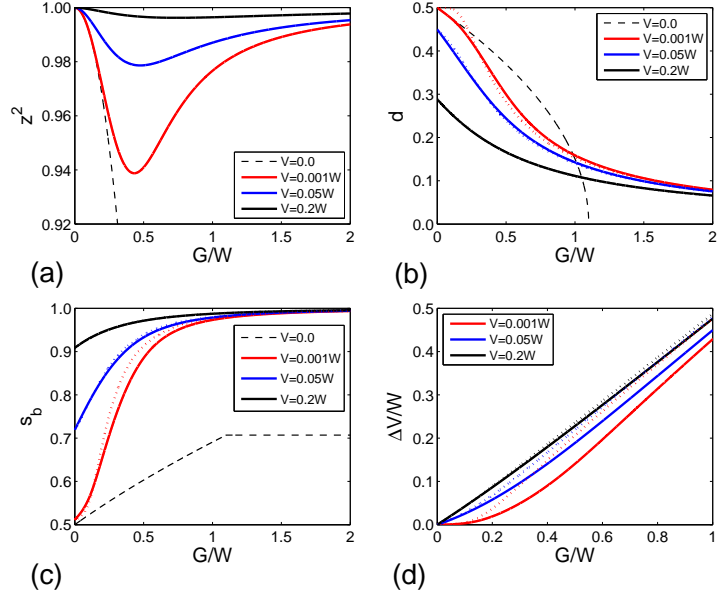} 
\caption{\label{sbfields_Vne0efeq0} (color online) Variation of the slave
  boson (SB)
  parameters with $G$ for $t_{ff}=-0.1$, $\epsilon_{f}=0$ and various values of
  $V$. In (b-d) the solid lines indicate the SB solution while the
  dotted lines indicate the Hartree-Fock solution. (a) Band renormalization
  factor $z^2$; (b) SB field $d$; (c) SB field $s_{b}$; (d)
  change in hybridization $\Delta{V}$.}   
\end{figure}

As shown in~\fig{sbfields_Vne0efeq0}(a), for $\epsilon_{f}=0$ and any finite
$V$ the band-renormalization factor $z^2$ decreases 
as $G$ is increased, but eventually goes through a minimum before asymptoting
to unity. 
For $V\ll{W}$, $z^2$ closely follows the band renormalization in the $V=0$
system until $G\approx0.2W$ where the curvature changes and a minimum
value is subsequently reached at $G\approx0.4W$. This has been checked down to
$V=10^{-5}W$ (not shown), which is found to give almost identical results
to the system with $V=10^{-3}W$. This has an important
implication: we do not recover the results of~\Sec{subsec:Veq0} in the limit
$V\rightarrow0$, indicating that there is a spontaneous hybridization in the
$V=0$ system. We will
discuss this EI phase in more detail in~\Sec{sss:EI}.
Although the minimum in $z^2$ suggests a crossover between two
distinct regimes, this interpretation has to be used with caution: 
since the depth of the minimum decreases with increasing $V$, the difference
between the low- and high-$G$ regimes becomes less pronounced.

To understand the nature of the low- and high-$G$ regimes we compare
in~\fig{sbfields_Vne0efeq0}(b-d) the SB 
and HF results for $d$, $s_b$ and the excitonic enhancement of the 
hybridization $\Delta{V}=\widetilde{V}-V$. For small $V$, the SB results
differ  
considerably from the HF results for $G\lesssim1$, although with increasing
$G$ the results of the two theories converge, which indicates that the 
physics of the high-$G$ regime is HF-like. This is consistent with the
$G\rightarrow\infty$ limit of $z^2$, which is unity as in a HF theory. 
This HF-like behaviour is observed for all $G$ at sufficiently large $V$: for
$V=0.2W$, the SB and HF results for $s_b$ and $d$ are almost identical for all
$G$, 
although there is still a noticeable difference in the $G$-dependence of
$\Delta{V}$. This is consistent with the very shallow minimum in $z^2$
observed in~\fig{sbfields_Vne0efeq0}(a).

For $G\lesssim0.2W$ and $V=0.001W$ the SB result for $d$ decreases linearly
with $G$, closely following the $V=0$ results. This is in contrast to the HF
theory, which predicts that $d$ differs from its non-interacting value by
$\Delta\sim\exp(-W/G)$ in this regime. The value of $s_b$, however, does 
not follow the $V=0$ results over the same range: this is due to the greater
sensitivity of $s_b$ to the magnitude of the hybridization, as can be seen by
examining the evolution of the $G=0$ values as $V$ is decreased. Further
lowering $V$ we indeed find that the $V=0$ results are tracked in the low-$G$
regime (not shown). These results, along with the variation of $z^2$, clearly
indicate the importance of correlations beyond HF theory at small
$G$ and $V$.

The different treatment of the on-site Coulomb interaction in the two
theories is essential to understanding the divergence between the HF and SB
results. In  
the HF theory, the Coulomb interaction indirectly affects the concentration of
doubly-occupied sites ($d^2$) through the
renormalization of the band parameters, in this case only the hybridization
(at $\epsilon_{f}=0$ the renormalization of the orbital energies is
identical). In the SB 
theory, however, the penalty for double occupancy of 
a site is also explicitly taken into account in the expression for the energy
by the  
$Gd^2$ term. Thus, for $G\ll{W}$ when the Coulomb renormalization of the
hybridization is exponentially small in both theories, the SB theory
nevertheless predicts a reduction $\propto{G^2}$ in $d^2$ whereas the HF
theory has only $\sim\exp(-2W/G)$  
reduction. At $G\gtrsim{W}$, where the renormalization of the hybridization is
large and grows linearly with $G$ for both theories, the $G$-dependence of the
SB and HF predictions for $d$ is almost equal. The interesting conclusion can
then be drawn that within the SB theory the on-site Coulomb repulsion is
less important to the physics than the renormalization of $V$ in the high-$G$
regime. That is, the excitonic enhancement of the hybridization gap
compensates for the energy penalty due to the finite concentration of
doubly-occupied sites. 

\begin{figure}
\includegraphics[width=9.5cm]{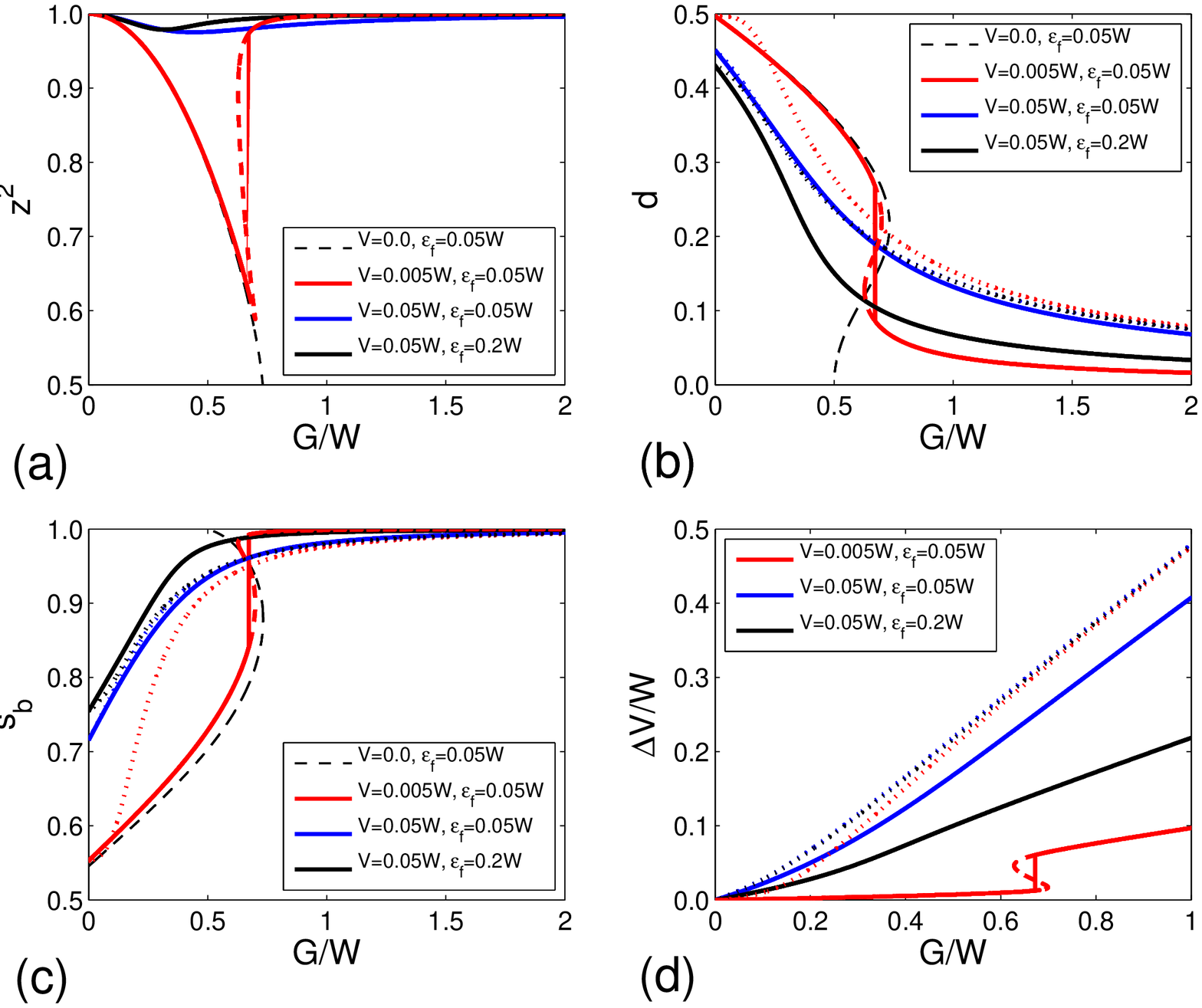} 
\caption{\label{sbfields_tlt0Vne0efne0} (color online) Variation of the slave
  boson (SB)
  parameters with $G$ and $V$ for $t_{ff}=-0.1$ and $\epsilon_{f}\neq0$. In
  all plots the dashed lines indicate unstable solutions; in
  (b-d) the solid lines indicate the SB solution while the 
  dotted lines indicate the Hartree-Fock solution. (a) Band renormalization factor
  $z^2$; (b) SB field $d$; (c) SB field $s_{b}$; (d)
  change in hybridization $\Delta{V}$.}   
\end{figure}

The $G$-dependence of the SB fields for $\epsilon_f\neq0$ is qualitatively
similar to the $\epsilon_{f}=0$ results for sufficiently large $V$, but as $V$
is reduced they 
converge towards the $V=0$ results of~\Sec{subsec:Veq0}. This can clearly be
seen for the $V=0.005W$, $\epsilon_f=0.05W$ line
in~\fig{sbfields_tlt0Vne0efne0}, which closely 
follows the $V=0$ curve almost until the valence transition is reached
at $G\approx0.7W$. Although $d$ does not vanish at higher $G$ when $V\neq0$,
it is heavily 
suppressed below its value in the HF theory. Intermediate between the low-
and high-$G$ regimes there is a small range of $G$ values for which two
solutions exist, implying a discontinuous evolution from the low- to the
high-$G$ regimes. Unlike the $\epsilon_{f}=0$ case
[\fig{sbfields_Vne0efeq0}(c)], for $G\lesssim0.7W$ the 
$G$-dependence of $s_b$ is very similar to that of the $V=0$ results [dashed
line in~\fig{sbfields_tlt0Vne0efne0}(c)]. The
renormalization of the  
hybridization remains small in the low-$G$ regime, but it is much larger
and grows linearly with $G$ beyond the jump discontinuity. This linear
increase is nevertheless slower than in the HF theory for all $\epsilon_{f}$
and $V$. This can be explained by the $V$-dependence of the
hybridization enhancement  
\beq
\Delta{V} = \frac{|V|}{\sqrt{4V^2+\epsilon_{f}^2}}(\Lambda_{a}-\Lambda_b).
\label{eq:deltaV} 
\eeq
In the high-$G$ regime the difference
$\Lambda_a-\Lambda_b$ grows linearly with the interaction strength. Replacing
the $V$-dependent prefactor by $0.5$ we then obtain close 
correspondence with the HF results, as
in~\fig{sbfields_Vne0efeq0}(d). The prefactor in~\eq{eq:deltaV}, however,
arises from   
the transformation to the $a$-$b$ basis and has important consequences for
the SB theory of the EFKM. In particular, for finite
$\epsilon_{f}$ the prefactor vanishes as $V\rightarrow0$, whereas
$\Lambda_a-\Lambda_b$ remains finite. This result implies that within the SB
theory there is no spontaneous hybridization  
if $\epsilon_{f}\neq0$, which is in strong contradiction to the HF
theory where the EI phase is stable at $T=0$ for
$|\epsilon_{f}|<(1+t_{ff})W/2$.~\cite{Khomskii,POS96} Indeed, for
$\epsilon_{f}\neq0$, the SB and HF theories are only in agreement when
$V\gg|\epsilon_{f}|$. 

\begin{figure}
\includegraphics[width=9.5cm]{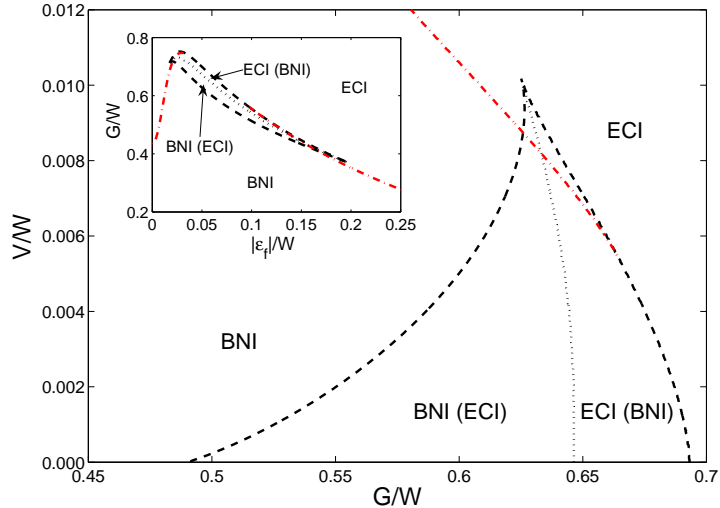} 
\caption{\label{PD_tflt0Vne0} (color online) Phase diagram in the
  $G$-$V$ plane for $t_{ff}=-0.1$ and $\epsilon_f=0.06W$. At high $G$ we have
  an excitonically-correlated insulator (ECI), whereas at low $G$ the
  band-narrowed insulator (BNI) is realized. The dashed lines
  bound the region where both BNI and ECI solutions are found, with the
  metastable state written in brackets. Along the dotted line the BNI and ECI
  solutions have equal energy. 
  The red dot-dash line indicates the location of the minimum in $z^2$: within
  the region where both BNI and ECI solutions are found this refers only to
  the BNI regime.
  The inset shows the phase diagram in the $G$-$\epsilon_f$ plane for
  $t_{ff}=-0.1$ and $V=0.005W$. Lines are the same as in the main figure.}   
\end{figure} 

The evolution of the model with decreasing $V$ is shown in the phase
diagram~\fig{PD_tflt0Vne0}. We classify the low- and high-$G$
regimes as the 
band-narrowed insulator (BNI) and the excitonically-correlated insulator
(ECI) respectively. The former reflects the reduced $z^2<1$ which is
characteristic of the $G<W$ solution when $V\ll{0.1W}$, whereas the
latter is due to the substantial excitonic renormalization of
the hybridization in the high-$G$ regime with characteristic
$\widetilde{V}\propto{G}$ dependence. These designations are most useful where
the 
evolution from the BNI to the ECI with increasing $G$ is discontinuous. When
there is no such discontinuity separating the low- and high-$G$ regimes, the
minimum in $z^2$ can serve as an approximate boundary. We emphasize that the
BNI and
ECI are not distinct \emph{phases} of the EFKM, as both are insulators and
there is no order parameter to distinguish between the two. Rather, they 
should be regarded as labeling regions of the 
phase diagram according to the dominant effect of the correlations. 
We see from the inset that at $V=0.005W$ multiple solutions are possible
within the thin region enclosed by the dashed lines, with the metastable state
indicated in brackets. The inset in~\fig{PD_tflt0Vne0} can be
directly compared
to~\fig{PD_Veq0}. As seen in the main figure, the region of multiple solutions
expands to fit the 
$V=0$ boundaries as $V$ is decreased, with the BNI (ECI) regimes evolving into
the M (FB) phase in the $V=0$ limit. 
Note the sensitivity of the region of multiple solutions to a
finite $V$: this vanishes completely for $V\gtrsim0.01W$. We conclude that
$V\neq0$ strongly suppresses any tendency to phase separation in the EFKM.

\subsection{$V\neq0$, $t_{ff}>0$} \label{ss:Vne0tgt0}

\begin{figure}
\includegraphics[width=9.5cm]{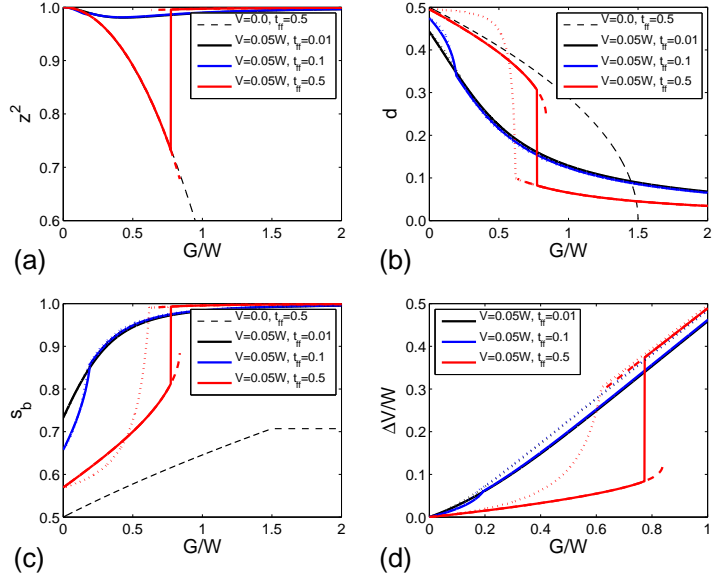} 
\caption{\label{sbfields_tgt0Vne0efeq0} (color online) Variation of the slave
  boson (SB)
  parameters with $G$ and $t_{ff}$ for $V=0.05$ and $\epsilon_{f}=0$. In
  all plots the dashed lines indicate unstable solutions; in
  (b-d) the solid lines indicate the SB solution while the 
  dotted lines indicate the Hartree-Fock solution. (a) Band renormalization
  factor 
  $z^2$; (b) SB field $d$; (c) SB field $s_{b}$; (d)
  change in hybridization $\Delta{V}$.}   
\end{figure}

The behaviour of the EFKM with $V\neq0$ and $t_{ff}>0$ is qualitatively
different 
to that for $t_{ff}\leq0$. For sufficiently small $V$ the $t_{ff}>0$
non-interacting ground state
is metallic, as shown in~\fig{bands}(d). As $G$ is increased,
however, the excitonic renormalization of the hybridization eventually opens a
gap and the system is then in the ECI phase.
In~\fig{sbfields_tgt0Vne0efeq0} we present representative examples of
the different behaviour displayed by the EFKM at constant
$V=0.05W$ and $\epsilon_{f}=0$. The behaviour of the system for
$\epsilon_{f}\neq0$ is qualitatively identical, although the critical coupling
for the metal-insulator transition (MIT) in the SB theory is greatly increased
at small $V$ due to the much smaller excitonic renormalization of the
hybridization than at $\epsilon_{f}=0$.   

The $t_{ff}=0.01$ data illustrates the behaviour of the EFKM when the
non-interacting system is insulating. The system remains in the ECI state for
all $G$, and there are no significant differences between
these results and those presented in~\Sec{ss:Vne0tle0}. More interesting is
the case $t_{ff}=0.1$ as there is a second-order MIT 
at $G\approx0.2W$. This is reflected in the abrupt change in the first
derivative 
with respect to $G$ of the curves in~\fig{sbfields_tgt0Vne0efeq0}. 
Because the MIT is driven by the excitonic
renormalization of the hybridization, which follows closely the HF values in
the insulating state, the SB and HF results for the critical coupling agree
very well. Since the 
effective hybridization in the SB theory is always smaller than in the HF
theory, however, the MIT occurs at a
slightly higher value of $G$ in the SB theory.

As $t_{ff}$ is further increased the MIT in the SB theory becomes
first order. This case is 
represented in~\fig{sbfields_tgt0Vne0efeq0} by the $t_{ff}=0.5$ results.
Note that there is a region around $G=0.7W$ where both the ECI and metallic (M)
phases are solutions to~\eq{eq:saddle_point}. Within the ECI phase the SB and
HF results are in close agreement for $d$, $s_b$ and $\Delta{V}$; in the M
phase, however, the two theories give very different predictions. This is due
to the much 
smaller effective hybridization within the SB theory
[see~\fig{sbfields_tgt0Vne0efeq0}(d)] as well as the importance of the
strong correlations in the M phase. The latter aspect is clearly shown
in~\fig{sbfields_tgt0Vne0efeq0}(a,b) 
by the close correspondence between $z^2$ and $d$ in the $V=0.05W$ and $V=0$
systems. Again, this is not seen in the $G$-dependence of $s_{b}$
due to the much greater sensitivity of $s_b$ to the value of $V$. Note that as
we lower $G$ from the ECI phase and enter into the metastable regime, the band
gap continuously vanishes as we approach the limit of
metastability. Increasing $G$ from within the M state, however, the band
overlap does not go to zero as the metastable limit is approached. 

\begin{figure}
\includegraphics[width=9.5cm]{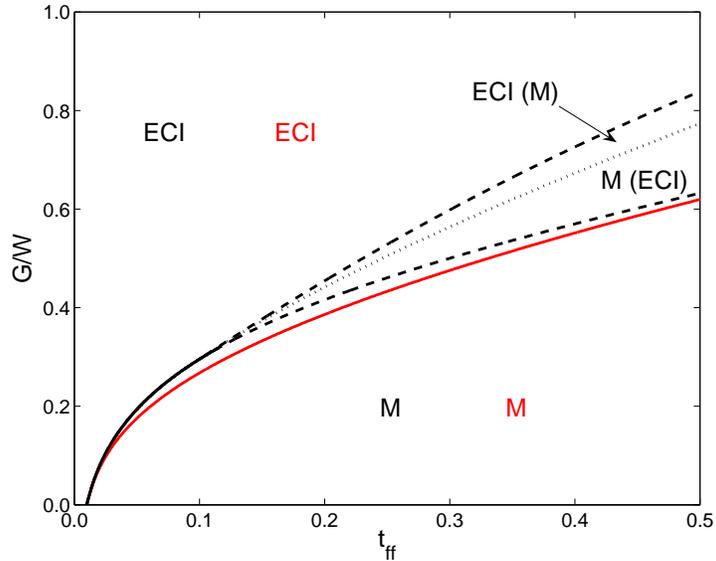} 
\caption{\label{PD_tfgt0Vne0} (color online) Phase diagram in the
  $G$-$t_{ff}$ plane for $V=0.05W$ and $\epsilon_f=0.0$. The red line and
  labels refer to the Hartree-Fock phase diagram, whereas the black lines and
  labels 
  refer to the slave boson theory. The solid line refers to a second-order
  transitions, 
  the dashed lines to the limits of metastability of the
  excitonically-correlated insulator (ECI) and metal (M) phases. The
  ECI and M phases have equal energy along the dotted line.}   
\end{figure} 

We present the phase diagram for the EFKM with $t_{ff}<0$ at $V=0.05W$ and
$\epsilon_{f}=0$ in~\fig{PD_tfgt0Vne0}. The phase boundaries in both
the SB and HF theories are included, the former in black and the latter in
red; the colour of the labels also refers to the different theories. In the HF
theory we find only a line of second-order transitions. For the SB theory the
MIT is of second order for $t_{ff}\lesssim0.11$ and is of first order at higher
values of $t_{ff}$, where a region with both M and ECI 
solutions is found. As in the previous
phase diagrams the metastable phase is given in brackets. 
The origin of the first-order MIT in the SB theory is related to the behaviour
of the EFKM in the limit $t_{ff}=1$. Here the SB solution of the EFKM with
$V\neq0$ is identical to the paramagnetic SB solution of the Hubbard model in a
transverse magnetic field, so for sufficiently small $V$ there is a first-order
transition into the FB state and a region of both M and FB solutions. As
$t_{ff}$ is reduced from unity, this region narrows until the lines of
metastability converge at $t_{ff}\approx0.11$. 

\subsection{Excitonic Insulator} \label{sss:EI}

\begin{figure}
\includegraphics[width=9.5cm]{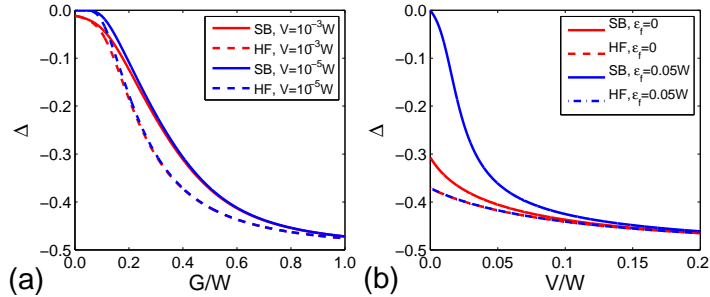} 
\caption{\label{EIDelta} (color online) (a) Scaling of
  the excitonic average $\Delta$ with $V$ for the Hartree-Fock (HF) and slave
  boson (SB) theories at
  $t_{ff}=-0.1$ and $\epsilon_f=0$. (b) $V$-dependence of the excitonic
  average $\Delta$ for the HF and SB theories at $t_{ff}=-0.1$ and $G=0.4W$.}     
\end{figure}

As noted in~\Sec{ss:Vne0tle0} the SB theory only allows an EI phase at
$\epsilon_{f}=0$, whereas the EI phase is a ubiquitous feature of the HF
solution. As we see in~\fig{EIDelta}(a), at $\epsilon_{f}=0$ the two theories
give good agreement for the excitonic average $\Delta$ when $G>W$ and
$V>0.1W$; for $G<W$ and $V\ll0.1W$, 
the SB results 
predict a smaller $|\Delta|$ as expected from the much smaller effective
hybridization shown in~\fig{sbfields_Vne0efeq0}(d). The predictions of the HF and
SB results diverge considerably for $\epsilon_{f}\neq0$ as displayed
in~\fig{EIDelta}(b). Note that in the HF theory $\Delta$ is almost identical
for $\epsilon_{f}=0$ and $\epsilon_{f}=0.05W$.
The very different results of the HF and SB mean-field approaches suggest
separate interpretations of the EI phase. 

In the HF theory the EI phase arises from the formation of an excitonic
condensate due to the attraction between $f$ holes and $c$ electrons. This
excitonic pairing creates an effective hybridization between the two bands, as
an electron can hybridize from a $c$ orbital into a $f$ orbital via the
formation and dissociation of an exciton. At $T=0$ the normal (N) state is
unstable to the EI phase if the effective hybridization is sufficient to open
a gap in the system. In particular, the EFKM with $t_{ff}\leq0$ has an
instability towards the EI phase for arbitrarily small inter-orbital Coulomb
repulsion $G$. For $t_{ff}>0$, however, the EI phase is only realized when the
effective hybridization is large enough to eliminate the band overlap and so
the N phase is stable up to a finite critical interaction strength.

To understand the results of the SB theory, we note that the $a$-$b$ basis
introduced in~\eq{eq:abbasis} rewrites the Hamiltonian in terms of bonding and 
anti-bonding orbitals at each site. 
The EI phase occurs when these orbitals have mixed $c$- and $f$-character
in the $V\rightarrow0$ limit. In the non-interacting system this condition is
only satisfied for $\epsilon_{f}=0$, when the transformation to the $a$-$b$
basis is independent of $V$, see~\eq{eq:abbasis}. For $\epsilon_{f}\neq0$, in
contrast, the transformation to the $a$-$b$ basis is $V$-dependent and the
bonding and anti-bonding orbitals continuously evolve into the $V=0$ atomic
orbitals as $V\rightarrow0$.
Due to the mixture of the two atomic orbitals in the $a$-$b$ basis, at finite
$G$ the Coulomb-induced splitting of the bonding and anti-bonding orbitals
produces the enhancement of the 
hybridization given by~\eq{eq:deltaV}. This induced hybridization only
survives in the $V\rightarrow0$ limit when $\epsilon_{f}=0$, as it is only in
this case that the atomic orbitals are equally mixed in the $a$-$b$ basis for
all $V$. As such, we obtain a spontaneous hybridization when $\epsilon_{f}=0$,
whereas for $\epsilon_{f}\neq0$ the system continuously evolves into the $V=0$
system as clearly evidenced by the phase diagram~\fig{PD_tflt0Vne0}.

\begin{figure}
\includegraphics[width=9.5cm]{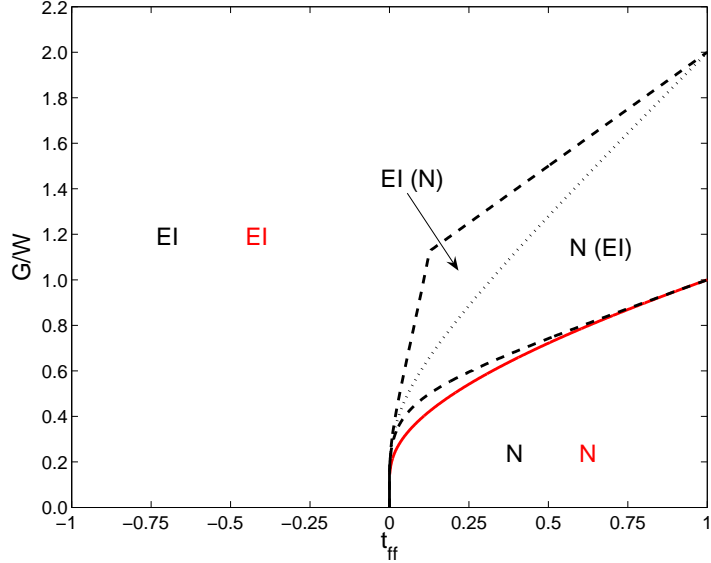} 
\caption{\label{PD_EI} (color online) $V=0$ phase diagram in the 
  $G$-$t_{ff}$ plane for $\epsilon_f=0.0$. The red
  line indicates the metal-insulator transition (MIT) between the excitonic
  insulator (EI) and normal (N) phases in Hartree-Fock theory, which 
  are indicated by the red labels. The slave-boson theory predicts a
  first-order MIT between the EI and N phases, with the boundaries
  of metastability 
  given by the dashed lines. The N and EI phases have equal energy along the
  dotted line.}
\end{figure}

The $\epsilon_{f}=0$ ground state phase diagram of the EFKM for 
$-1\leq{t_{ff}}\leq1$ is presented in~\fig{PD_EI}. Again, we plot the phase
diagram for the HF and SB theories in red and black respectively. For the HF
theory the second-order MIT at $t_{ff}>0$ found in~\Sec{ss:Vne0tgt0}
remains. Although the band gap vanishes continuously as we approach the
metallic state from within the EI phase, the excitonic average $\Delta$
discontinuously 
drops to zero at this line and so below it the system is in the N state.
In contrast to the HF results, the SB theory predicts a first-order MIT for
all $t_{ff}>0$ in the limit of vanishing hybridization. As
in~\Sec{ss:Vne0tgt0}, the lower limit of metastability of the EI phase lies
just above the MIT in the HF theory, and the band gap continuously vanishes
as the line of metastability is
approached, whereas $\Delta$ tends to a finite value. A finite band overlap
is always found as we approach the upper 
limit of metastability from within the N phase. Of particular note is the
$t_{ff}$-dependence of this line: for
sufficiently large $t_{ff}$ the boundary is given by $G=(1+t_{ff})W$, which
corresponds to the limit of metastability of the metallic phase at
$\epsilon_{f}=0^{+}$ found in~\Sec{subsec:Veq0}. At $t_{ff}\approx0.1$,
however, there is an abrupt change in this line, with the limit of
metastability of the N phase converging to $G=0$ at $t_{ff}=0$. 

\section{Conclusions} \label{sec:concl}

In this work we have studied the EFKM within the SB mean-field theory
introduced by Kotliar and Ruckenstein.~\cite{KR86} For the system with
$V\neq0$ we have 
compared the predictions of SB theory to those of the standard HF
approach.~\cite{Leder78} We have found that for the EFKM with $V=0$ the SB
phase diagram displays strong similarities to the SB solution of the
paramagnetic Hubbard model. In particular, at $\epsilon_{f}=0$ we find a
transition into a Brinkman-Rice-like insulating state as $G$ is increased,
whereas at small finite $\epsilon_{f}$ there is a first-order valence 
transition. For 
finite $V$ a distinction between the EFKM with $t_{ff}\leq0$ and $t_{ff}>0$
must be made. In the former case, the 
model is always in an insulating state and there is a considerable
renormalization of the hybridization by the Coulomb interaction. For
$\epsilon_{f}=0$ the SB and HF results are in good agreement, whereas for
$\epsilon_{f}\neq0$ the SB and HF theories only coincide when
$V\gg|\epsilon_{f}|$. For $V\ll|\epsilon_{f}|$ the SB theory displays a
crossover between a state with strong correlations beyond HF-level at
$G\ll{W}$ and an 
excitonically-correlated state at $G\gtrsim{W}$.
At sufficiently large $t_{ff}>0$ both the HF and SB theories predict a MIT as
$G$ is increased. In the former this is always of second order; for the latter
the MIT is of first order for $t_{ff}$ greater than some critical value. The
presence of first-order transitions in our SB treatment suggests that it is
worthwhile to include the inter-orbital Coulomb repulsion as an important
factor in discontinuous valence transitions.

We have also studied the appearance of an EI phase within the EFKM and our
conclusions severely constrain the parameter space of the model where such a
state is possible. In contradiction to the results of HF theory where the EI
phase is a ubiquitous feature of the $T=0$ phase
diagram,~\cite{Khomskii,POS96} the SB 
theory only predicts a spontaneous hybridization when the $c$ and
$f$ electron atomic orbitals are degenerate, i.e. $\epsilon_{f}=0$. When
$\epsilon_{f}\neq0$, the effective hybridization continuously vanishes as
$V\rightarrow0$. We have explained this difference in terms of the importance
of the bonding and anti-bonding orbitals in the SB theory. This imposes a
condition on the atomic structure for the realization of the EI phase. Such a
condition is absent in the HF theory. 

Our work has only considered uniform ground states of the EFKM. From the
rigorous solution of the FKM it is expected that the EFKM on a bipartite
lattice has an instability towards a density-wave state at
$\epsilon_{f}=0$.~\cite{BinaryAlloy} As demonstrated by a number of authors,
this density-wave state is stable at $T=0$ and prevents the EI phase from
being realized.~\cite{C99,BZGB05,FarkyDW} It is likely that this density-wave
phase would also be found within a SB treatment of the model. As the condition
for the EI phase in the SB theory concerns only the atomic as
opposed to the band structure, however, the EI phase could still be realized
in systems where the mean-field density-wave state is unstable, e.g.
frustrated lattices. A further extension of our work would be to examine the
effect of doping away from half-filling. 

\section*{Acknowledgments}

This work was funded by the European Union CoMePhS project. The author
gratefully acknowledges H. Yamase and M. Gul\'{a}csi for many helpful
discussions and their critical reading of the manuscript. K. I. Kugel is also
thanked for useful discussions.

\end{document}